\documentclass{llncs}
\usepackage{amsmath,amssymb,amsfonts,epsfig,algorithm,algorithmic,url,mathtools}
\usepackage{array,multirow,hyperref,breakcites,dsfont,graphicx}
\usepackage[inline]{enumitem}
\hypersetup{
     colorlinks   = true,
     citecolor     = blue,
     urlcolor     = blue
}
\usepackage{multirow}
\usepackage{graphicx}	
\usepackage{booktabs,threeparttable}
\newcommand{\norm}[1]{\left\lVert#1\right\rVert}
\begin{document}
\title{Grade Prediction with Course and Student Specific Models}
\institute{University of Minnesota}
\author{Agoritsa Polyzou \and George Karypis}
\institute{University of Minnesota}
\maketitle	

\begin{abstract}

The accurate estimation of students' grades in future courses is important as it can
inform the selection of next term's courses and create personalized degree pathways
to facilitate successful and timely graduation.
This paper presents future-course grade predictions methods based on sparse linear
models and low-rank matrix factorizations that are specific to each course or
student-course tuple. These methods identify the predictive subsets of prior courses
on a course-by-course basis and better address problems associated with the
\emph{not-missing-at-random} nature of the student-course historical grade data. 
The methods were evaluated on a dataset obtained from the University of Minnesota.
This evaluation showed that the course specific models outperformed various competing
schemes with the best performing scheme achieving a RMSE across the
different courses of 0.632 vs 0.661 for the best competing method.

\end{abstract}

\section{Introduction}

Data mining and machine learning approaches are being increasingly used to analyze educational- and learning-related datasets towards understanding how students learn and improving learning outcomes. This has led to the development of various approaches for modeling and predicting the success or failure of students in completing specific tasks in the context of intelligent tutoring systems~\cite{romero2008,toscher2010,pardos2010,thai2012,mckay2012,hwang2015}, building intelligent ``early warning systems'' that monitor the students' performance during the term~\cite{starfish,arnold2012}, predicting how well the students will perform by analyzing their activities with the learning management system (e.g., Moodle)~\cite{Elbadrawy2015a,sorour2015,luo2015}, and predicting students' term and final GPA~\cite{osman2012,ogu2014,al2016}. 

Our work focuses on developing methods that utilize historical student-course grade information to accurately estimate how well students will perform (as measured by their grade) on courses that they have not yet taken. Being able to accurately estimate students' grades in future courses is important as it can be used by them (and/or their academic advisers) to identify the appropriate set of courses to take during the next term, and create personalized degree pathways that enable them to successfully and effectively acquire the required knowledge to complete their studies in a timely fashion.

Existing approaches for predicting a student's grade in a future course~\cite{compass,denley2012,bydzo2015} rely on neighborhood-based collaborative filtering methods~\cite{kantor2011}. 
Despite their relative simplicity, the estimations obtained by these methods are reasonably accurate indicating that there is sufficient information in the historical student-course grade data to make the estimation problem feasible.

In this paper we improve upon these methods by developing various future-course grade prediction methods that utilize approaches based on sparse linear models and low-rank matrix factorizations. These methods rely entirely on the performance that the students achieved in previously taken courses.  A unique aspect of many of our methods is that their associated models are either specific to each course or specific to each student-course tuple. This allows them to identify and utilize the relevant information from the prior courses that are associated with the grade for each course and better address problems associated with the \emph{not-missing-at-random} nature of the student-course historical grade data. We experimentally evaluated the performance of our methods on a dataset obtained from the University of Minnesota that contained historical grades that span 12.5 years. Our results showed that the course specific models outperformed various competing schemes and that the best performing scheme, which is based on course-specific regression, achieves a RMSE across the different courses of 0.632 whereas the best competing method achieves an RMSE of 0.661.

The reminder of the paper is organized as follows. Section~\ref{sec:defnot} introduces the notation and definitions used. Section~\ref{sec:methods} describes the methods developed and section~\ref{sec:expinfo} provides information about the experimental design. Section~\ref{sec:results} presents an extensive experimental evaluation of the methods and compares them against existing approaches. Finally, Section~\ref{sec:concl} provides some concluding remarks.

\section{Definitions and Notations}
\label{sec:defnot}

Throughout the paper, bold lowercase letters will denote column vectors (e.g., $\mathbf{y}$) and bold uppercase letters will denote matrices (e.g., $\mathbf{G}$). Individual elements will be denoted using subscripts (e.g., for a vector ${y}_i$, and for a matrix $g_{s,c}$). A single subscript on a matrix will denote its corresponding row. The sets will be represented by calligraphic letters. 

The historical student-course grade information will be represented by a sparse matrix $\mathbf{G} \in \mathbb{R}^{n \times m}$, where $n$ and $m$ are the number of students and courses, respectively, and $g_{i,j}$ is the grade in the range of [0,4] that student $i$ achieved in course $j$. If a student has not taken a course, the corresponding entry will be missing. The course and student whose grades need to be predicted will be called \emph{target course} and \emph{target student}, respectively. 

\section{Methods}
\label{sec:methods}

In this section we describe various classes of methods that we developed for predicting the grade that a student will obtain on a course that he/she has not yet taken.  

\subsection{Course-Specific Regression (CSR)}
\label{methods:csr}

Undergraduate degree programs are structured in such a way that courses taken by students provide the necessary knowledge and skills for them to do well in future courses. As a result, the performance that a student achieved in a subset of the earlier courses can be used to predict how well he/she will perform in future courses. Motivated by this, we developed a grade prediction method, called \emph{course-specific regression} (CSR) that predicts the grade that a student will achieve in a specific course as a sparse linear combination of the grades that the student obtained in past courses.

In order to estimate the CSR model for course $c$, we extract from the overall student-course matrix $\bf{G}$ the set of rows corresponding to the students that have taken $c$. For each of these students (rows), we keep only the grades that correspond to courses taken prior to course $c$. Let ${\bf G}^c  \in \mathbb{R}^{n_c \times m}$ be the matrix representing that extracted information, where $n_c$ is the number of students that took course $c$. In addition, let ${\bf{y}}^c \in \mathbb{R}^{n_c}$ be the grades that the students in ${\bf G}^c$ obtained in course $c$ (the ${y^c_i}$ is the grade corresponding the the student in the $i$th row of ${\bf G}^c$). Given this, the CSR model ${\bf w}^c \in \mathbb{R}^{m}_{+}$ for $c$ is estimated as:
\begin{equation}
\underset{{\bf w}^c \succeq 0}{\mbox{minimize}}~\norm{\mathbf{y}^c-\mathds{1}w^c_0-{\bf G}^c{\bf w}^c}^2_2 + \lambda_{1}\norm{{\bf w}^c}^2_2 + \lambda_{2}\norm{{\bf w}^c}_1, 
\label{eq:1}
\end{equation} 
where $w^c_0$ is a bias term, $\mathds{1} \in \mathbb{R}^{n_c}$ is a vector of ones and $\lambda_{1},\lambda_{2}$ are regularization parameters to control overfitting and promote sparsity. The model is non-negative because we assume that prior courses can only provide knowledge to future courses. The individual weights of ${\bf w}^c$ indicate how much each prior course contributes in the prediction and represent a measure of the importance of the prior course within the context of the estimated model. Using this model, the grade that a student will obtain in course $c$ is estimated as
\begin{equation}
{\hat{y}^c}= w^{c}_0+{\bf s}^T{\bf w}^c,
\label{eq:2}
\end{equation} 
where ${\bf s} \in \mathbb{R}^m$ is the vector of the student's grades in the courses he/she has taken so far.

We found that by centering each student's grades around his/hers GPA leads to more accurate predictions (see Section~\ref{sec:results:csr}). In this approach, prior to estimating the model using Equation~\ref{eq:1}, we first subtract from each $g^c_{i,j}$ grade the GPA of each student (GPA is calculated based on the information in ${\bf G}^c$). This centers the data for each student and takes into consideration a notion of student bias as it predicts the performance with respect to the current state of a student. 
Note that in the case of GPA-centered data, we remove the non-negativity constraint on ${\bf w}^c$. We will refer to this model as the CSR-RC (Row Centered) model.

\subsection {Student-Specific Regression (SSR)}
\label{methods:ssr}

Depending on the major, the structure of different undergraduate degree programs can be different. Some degree programs have limited flexibility as to the set of courses that a student has to take and at which point in their studies they can take them (i.e., specific semester). Other degree programs are considerably more flexible and are structured around a fairly small number of core courses and a large number of elective courses. 

For the latter type of degree programs, a drawback of the CSR method is that it requires the same linear regression model to be applied to all students. However, given that the set of prior courses taken by students in such flexible degree programs can be quite different, a single linear model can fail to capture the various prior course combinations. In fact, there can be cases in which many of the most important courses that were identified by the CSR model were simply not taken by some students, even though these students have acquired the necessary knowledge and skills by taking a different set of courses. To address this limitation, we developed a different method, called \emph{student-specific regression} (SSR),  which estimates course-specific linear regression models that are also specific to each student.

The student specific model is derived by creating a student-course specific grade matrix ${\bf G}^{s,c}$ for each target student $s$ and each target course $c$ from the ${\bf G}^c$ matrix used in CSR method. ${\bf G}^{s,c}$ is created in two steps. First, we eliminate from ${\bf G}^c$ any grades for courses that were not taken by the target student. Second, we eliminate from ${\bf G}^c$ the rows that correspond to students that have not taken a sufficient number of courses that are in common with the target student $s$. Specifically, if $\mathcal{C}_s$ and $\mathcal{C}_i$ are the set of courses for student $s$ and $i$ respectively, we compute the overlap ratio (OR)
$={|\mathcal{C}_s \cap \mathcal{C}_i|}/{|\mathcal{C}_s|}$
and if OR$<t$, then student $i$ is not included in ${\bf G}^{s,c}$. The value of $t$ is a parameter of the SSR method and high values ensure that the set of students forming ${\bf G}^{s,c}$ have taken many courses in common with $s$ and have followed similar degree plans. Given ${\bf G}^{s,c}$, the SSR method proceeds to estimate the model using Equation~\ref{eq:1} (with ${\bf G}^{s,c}$ replacing ${\bf G}^{c}$), and uses Equation~\ref{eq:2} for prediction.

\subsection{Methods based on Matrix Factorization}
\label{sec:methods:mf}

Low rank matrix factorization (MF) approaches have been shown to be very effective for accurately estimating ratings in the context of recommender systems~\cite{kantor2011}. These approaches can be directly applied to the problem of predicting the grade that a student will achieve on a particular course by treating the student-course grade matrix $\bf{G}$ as the user-item rating matrix. 

The use of such MF-based approaches for grade prediction is postulated on the fact that there is a low dimensional latent feature space that can jointly represent both students and courses. Given the nature of the domain, this latent space can correspond to the space of knowledge components. Each course vector is the set of components associated with a course and each student vector represents the student's level of knowledge across these knowledge components. 

By applying the common approaches of MF-based rating prediction to the problem of grade prediction, the grade that student $i$ will obtain on course $j$ is estimated to be
\begin{equation}
{\hat{g}_{i,j}}= \mu + sb_i + cb_j + {\mathbf{p}_i\mathbf{q}_j}^T,
\label{eq:3}
\end{equation} 
where $\mu$ is a global bias term, $sb_i$ and $cb_j$ are the student and course bias terms, respectively, and $p_i$ and $q_j$ are the latent representations for student $i$ and course $j$, respectively. The parameters of the MF method ($\mu, {\bf sb} \in \mathbb{R}^n, {\bf cb} \in \mathbb{R}^m, {\bf P} \in \mathbb{R}^{n \times l}$, and ${\bf Q} \in \mathbb{R}^{n \times l}$) are estimated following a matrix completion approach that considers only the observed entries in $\bf{G}$ as
\begin{equation}
\begin{split}
\underset{\mu, {\bf sb}, {\bf cb}, {\bf P}, {\bf Q}}{\mbox{minimize}} ~ \sum_{g_{i,j} \in {\bf G}}  {(g_{i,j}-\mu-sb_i-cb_j-  {\bf p}_i{\bf q}^{T}_{j})}^2 + \lambda & (\norm{{\bf P}}^2_F + \norm{{\bf Q}}^2_F + \\
& \norm{{\bf sb}}^2_2 + \norm{{\bf cb}}^2_2),
\end{split}
\label{eq:4}
\end{equation}
where $\lambda$ is a regularization parameter and $l$ is the dimensionality of the latent space, which is a parameter to this method.

The accurate recovery of the low rank model (when such a model exists) from a set of partial observations depends on having a sufficient number of observed entries, and on these entries be randomly sampled from the entries of the target matrix $\mathbf{G}$~\cite{chen2013}. However, in the context of student grade data, the set of courses that students take is not a random subset of the courses being offered as they need to satisfy their degree program requirements. As a result, such an MF approach may lead to suboptimal prediction performance.

In order to address this problem we developed a \emph{course specific matrix factorization} (CSMF) approach that estimates an MF model for each course by utilizing a course specific subset of the data that is denser (in terms of the number of observed entries and the dimensions of the matrix). As a result, it contains a larger number of random by sampled subsets of sufficient size.

Given a course $c$ and a set of students $\mathcal{S}^c$ for which we need to estimate their grade for $c$ (i.e., the students in $\mathcal{S}^c$ have not taken this course yet), the data that CSMF utilizes are the following: 
\begin{enumerate*}[label=(\roman*)]
  \item the students and grades of the $\mathbf{G}^c$ matrix and ${\bf y}^c$ vector of the CSR method (Section~\ref{methods:csr}),
  \item the students in $\mathcal{S}^c$ and their grades.
\end{enumerate*}
This data is used to form a matrix ${\bf X}^c \in \mathbb{R}^{(n_c+n_t) \times (m_c+1)}$, where $n_c$ is the number of students in $\mathbf{G}^c$, $n_t = |\mathcal{S}^c|$, and $m_c$ is the number of distinct courses that have at least one grade in $\mathbf{G}^c$ or $\mathcal{S}^c$. The values stored in ${\bf X}^c$ are the grades that exist in  $\mathbf{G}^c$ and $\mathcal{S}^c$. The last column of ${\bf X}^c$ stores the grades ${\bf y}^c$ for the course $c$ that were obtained from the students in $\mathbf{G}^c$. Thus, ${\bf X}^c$ contains all the prior grades associated with the students who have already taken course $c$ and the students for which we need to have their grade on $c$ predicted. Matrix ${\bf X}^c$ is then used in place of matrix $\bf G$ in Equation~\ref{eq:4} to estimate the parameters of the CSMF method, which are then used to predict the missing entries of the last column of ${\bf X}^c$, which are the grades that need to be predicted.

\section{Experimental Design}
\label{sec:expinfo}

\subsection{Dataset}
\label{sec:exp:datasets}

The student-course-grade dataset that we used in our experiments was obtained from the University of Minnesota which has a very flexible degree program. It contains the students that have been part of the Computer Science and Engineering (CSE) and Electrical and Computer Engineering (ECE) programs from Fall of 2002 to Spring of 2014. Both of these degree programs are part of the College of Science \& Engineering (CS\&E)  in which students have to take a common set of core science courses during the first 2--3 semesters. We removed from the dataset any courses that are not part of those offered by CS\&E departments, as these  correspond to various liberal arts and physical education courses, which are taken by few students and in general do not count towards degree requirements. Furthermore, we eliminated any courses that were taken as pass/fail.  The initial grades were in the A--F scale, which was converted to the 4--0 scale using the standard letter-grade to GPA conversion. The resulting dataset consists of 2,949 students, 2,556 different courses, and 76,748 student-course grades.  

We used this dataset to assess the performance of the different methods for the task of predicting the grades that the students will obtain in the last semester (i.e., the most recent semester for which we have data). For this reason, the dataset was further split into two parts, one containing the students that are still \emph{active}, i.e., have taken courses in the last semester ($D_{active}$) and one that contains the remaining students ($D_{inactive}$).  $D_{active}$ contains 876 students, 19,089 grades, out of which 3,427 grades are for the 475 distinct classes taken in the last semester. $D_{inactive}$ contains 2,073 students and 57,659 grades.

These datasets were used to derive various training and testing datasets for the different methods that we developed. Specifically, for the CSR  method we extracted the course specific training and testing datasets as follows. For each course $c$ that was offered in the last semester, we extracted course-specific training and testing sets ($D^{c, \ge k}_{train}$ and $D^{c, \ge k}_{test}$) by selecting from $D_{inactive}$ and $D_{active}$, respectively, the students that have taken $c$, and prior to taken $c$, they also took at least $k$ other courses. The reason that these datasets were parametrized with respect to $k$ is because we wanted to assess how the methods perform when different amount of historical student performance information is available. In our experiments we used $k$ in the set $\{5, 7, 9\}$. That information will create the grade matrix ${\bf G}^c$, where $g^{c}_{i,j}$ is the grade of the $i$th student on the $j$th course from the training set $D^{c, \ge k}_{train}$. Table~\ref{table:1} shows various statistics about the various course-specific datasets for different values of $k$. 

\begin{table}[bt]
\centering
\caption{Statistics for Course-Specific datasets.}
\setlength{\tabcolsep}{.5em}
\begin{tabular}{l |r r r}
\toprule
Prior courses		& 5  		& 7 		& 9 \\ 
\midrule
Average number of students in training set & 270 & 232 & 212\\
Average number of students in test set & 22 & 21 & 20\\
Average number of prior courses & 141 & 141 & 145\\
Average number of grades & 3,872 & 3,663 & 3,663\\
Courses predicted & 92 & 90 & 80\\
Grades predicted & 2,088 & 1,959 &1,666\\
\bottomrule
\end{tabular}
\label{table:1}
\end{table}

For the CSMF method, the training dataset for course $c$ was obtained by combining $D^{c, \ge k}_{train}$ and $D^{c, \ge k}_{test}$ into a single matrix after removing the grades that the target students achieved in course $c$.

For the MF method, the matrix is constructed as the union of the sets $D^{c, \ge k}_{train}$ and $D^{c, \ge k}_{test}$ for every course to be predicted after removing the grades that the active students achieved in the courses we want to predict. We formulated the dataset in this way in order to provide the same information for training and testing to all our models. 

\begin{figure}[bt]
    \centering
    \includegraphics[scale=0.35]{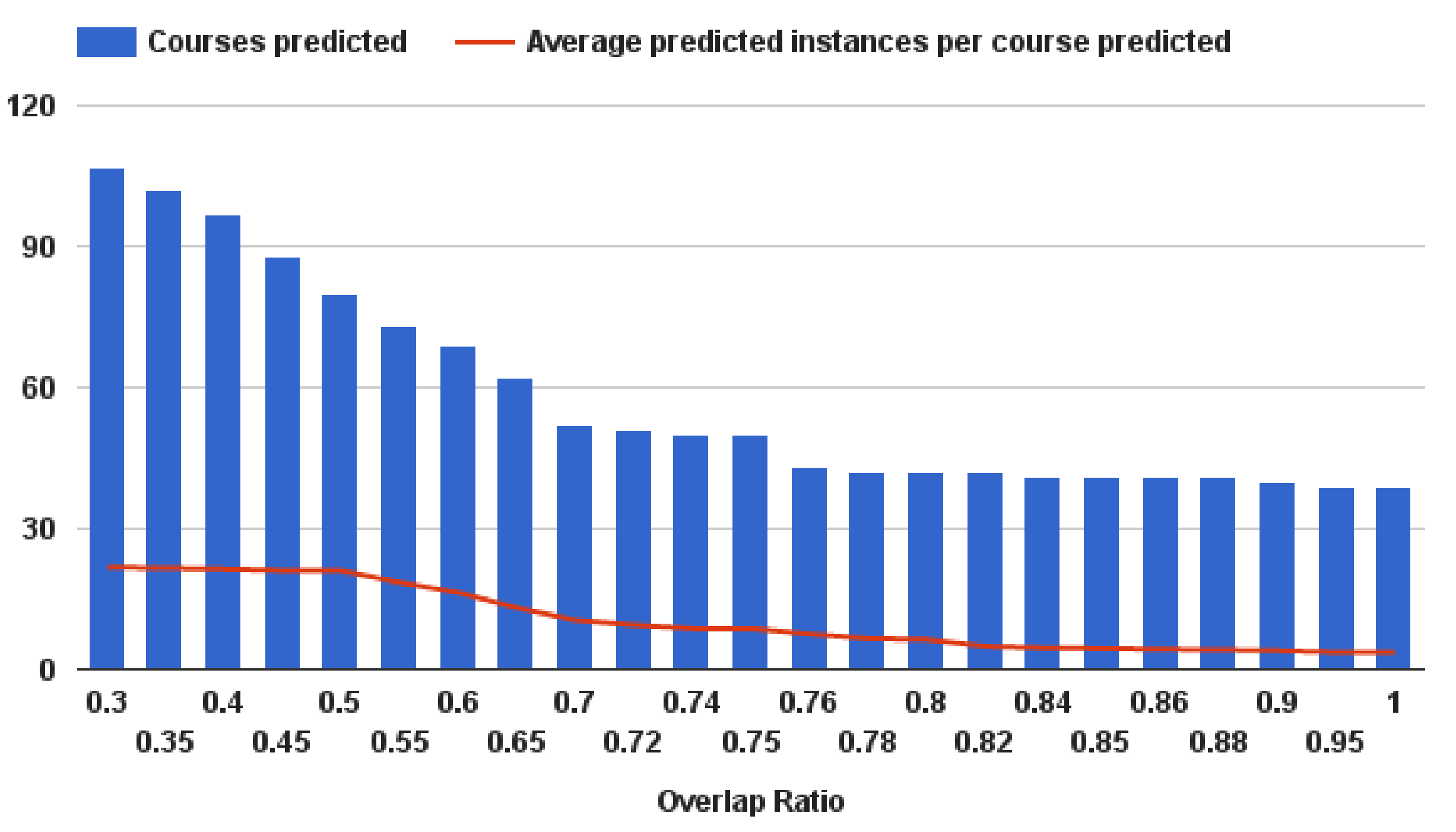}
    \caption{Statistics of the datasets used in SSR w.r.t overlap ratio.}
    \label{fig:1}
\end{figure}

In the SSR, the grade matrix $\mathbf{G}^{s,c}$ is created by selecting from $D^{c, \ge k}_{train}$ the set of courses that were also taken by student $s$ and the set of students whose OR with $s$ is at least $t$. Figure~\ref{fig:1} shows some statistics about these datasets as a function of $t$.

Finally, we did not consider the models that have less than 20 students in their corresponding dataset, as we consider them to have too few training instances for reliable estimation.

\subsection{Competing Methods}
\label{sec:methods:compm}

In our experiments, we compared our methods with the following competing approaches.
\begin{enumerate}
\item \textbf{BiasOnly.} We only took into consideration local and global bias to predict the students' grades. These biases were estimated using Eqn.~\ref{eq:4} when $l=0$.
\item \textbf{Student-Based Collaborative Filtering (SBCF).} This method implements the approach described in~\cite{bydzo2015}. For a target course $c$, every student $i$ is represented by a vector formed with his/hers grades in courses taken prior to $c$. The vector of a target student $s$ is compared against the vectors of the other students that have taken course $c$ with the Pearson's correlation coefficient. We select the students with positive similarity to perform grade prediction for $s$ in $c$ according to:
\begin{equation}
{\hat{g}_{s,c}}= \bar{g_s} + \frac{\min(r,nbr)}{r}\frac{\sum_{i=1}^{nbr}(g_{i,c}-\bar{g_i})sim_{s,i}}{\sum_{i=1}^{nbr}sim_{s,i}},
\label{eq:3}
\end{equation}
where $nbr$ is the number of students selected, $r$ is a confidence lower limit for significance weighting, $\bar{g_i}$ is the average grade of the student prior taking $c$, and $sim_{s,i}$ represents the similarity of target student $s$ with $i$.
\end{enumerate}

\subsection{Parameters and Model Selection}
For CSR, we let $\lambda_1$ take values from 0 to 40 in increments of 2.5 and $\lambda_2$ from 0 to 50 in increments of 2.5. For SSR, we let $\lambda_1$ take values from 0 to 10 in increments of 1 and $\lambda_2$ from 0 to 14 in increments of 2. For MF and CSMF, we let $\lambda$ take values from 0 to 6 in increments of 0.05. For SSR, the range of the tested values for overlap ratio is 0.3 to 1, in increments of 0.04. For MF and CSMF methods we tested the number of latent dimensions with the values 2, 5 and 8.

As we could not use cross validation for the SSR, we did not apply it for any regression model, in order to be fair with our comparisons. The best models are selected based on their performance on the test set. For MF based approaches, we used the semester before the target semester to estimate and select the best parameters. 

\subsection{Evaluation Methodology \& Performance Metrics}
We evaluated the performance of the different approaches by using them to predict the grades for the last semester in our dataset using the data from the previous semester for training.

We assessed the performance using the root mean square error (RMSE) between the actual grades and the predicted ones. Since the courses whose grades are predicted have different number of students, we computed two RMSE-based metrics. The first is the overall RMSE in which all the grades across the different courses were pooled together, and the second is the average RMSE obtained by averaging the RMSE values for each course. We will denote the first by RMSE and the second as AvgRMSE.

\section{Experimental Results}
\label{sec:results}

\subsection {Course-Specific Regression}
\label{sec:results:csr}

Table~\ref{table:2} shows the performance achieved by the CSR and CSR-RC models when trained using the three different datasets discussed in Section~\ref{sec:exp:datasets}. These results show that among the two models, CSR-RC, which operates on the GPA-centered grades leads to considerably lower errors both in terms of RMSE and AvgRMSE. In terms of the sensitivity of their performance  on the amount of historical information that was available when estimating these models (i.e., the minimum number of prior courses), we can see that for CSR-RC, the RMSE performance of the models does not change significantly; though the AvgRMSE performance improves when going from five to nine prior courses. This indicates that training sets with more number of prior courses tend to help smaller courses.

\begin{table}[bt]
\centering
\caption{The performance achieved by Linear Course-Specific Regression.}
\setlength{\tabcolsep}{.5em}
\begin{tabular}{l |r r r || r r r}
\toprule
	& \multicolumn{3}{c||}{RMSE} &\multicolumn{3}{c}{AvgRMSE}\\ \midrule
Prior courses		& 5  	& 7 	& 9  	& 5  	& 7 	& 9\\ 
\midrule
{CSR}			& 0.751 &0.761 &0.779 &0.757 &0.785 &0.762 \\
{CSR-RC}                & 0.634 &0.632 &0.632 &0.585 &0.579 &0.543 \\
\bottomrule
\end{tabular}
\label{table:2}
\begin{tablenotes}
	\item The performance of the models trained on the different datasets were evaluated on the $D_{test}^{\ge 9}$ test set, which is the common subset among their respective test sets.
\end{tablenotes}
\end{table}

\subsection{Student-Specific Regression}
\label{sec:results:ssr}

\begin{figure}[bt]
    \centering
    \includegraphics[scale=0.43]{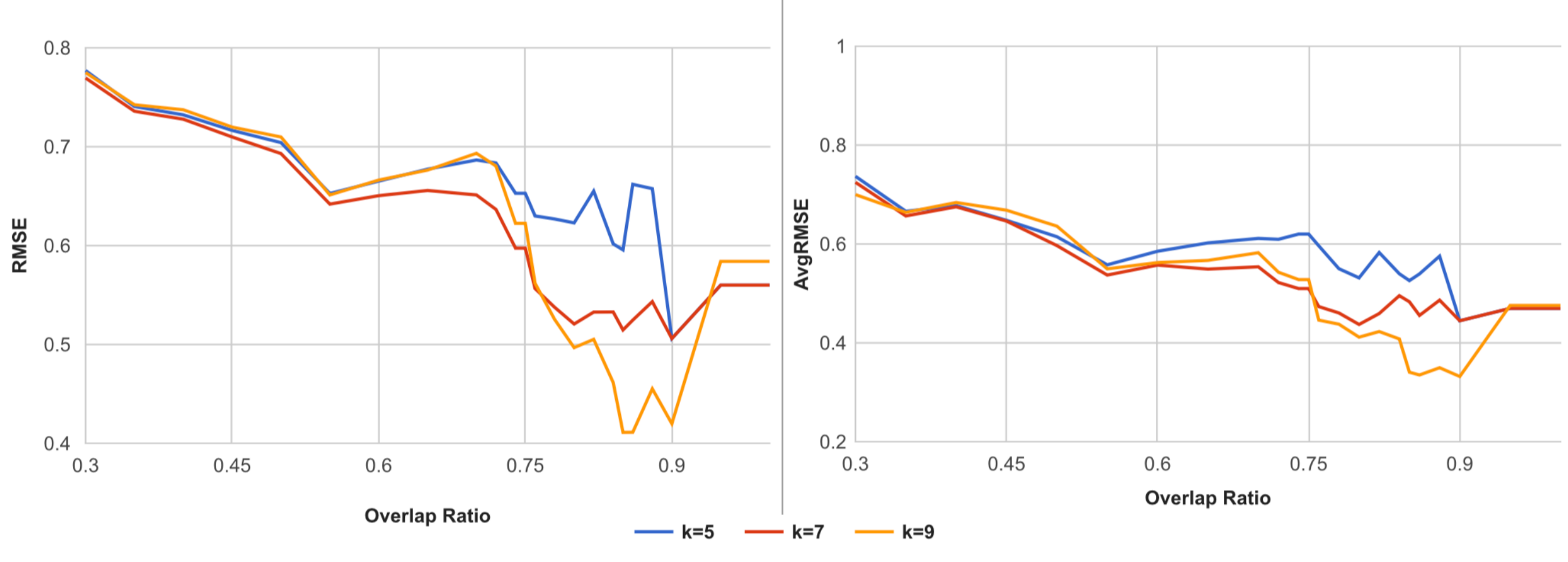}
    \caption{The performance achieved by the SSR model w.r.t. overlap ratio. }
    \label{fig:2}
\end{figure}

\begin{figure}[bt]
    \centering
    \includegraphics[scale=0.34]{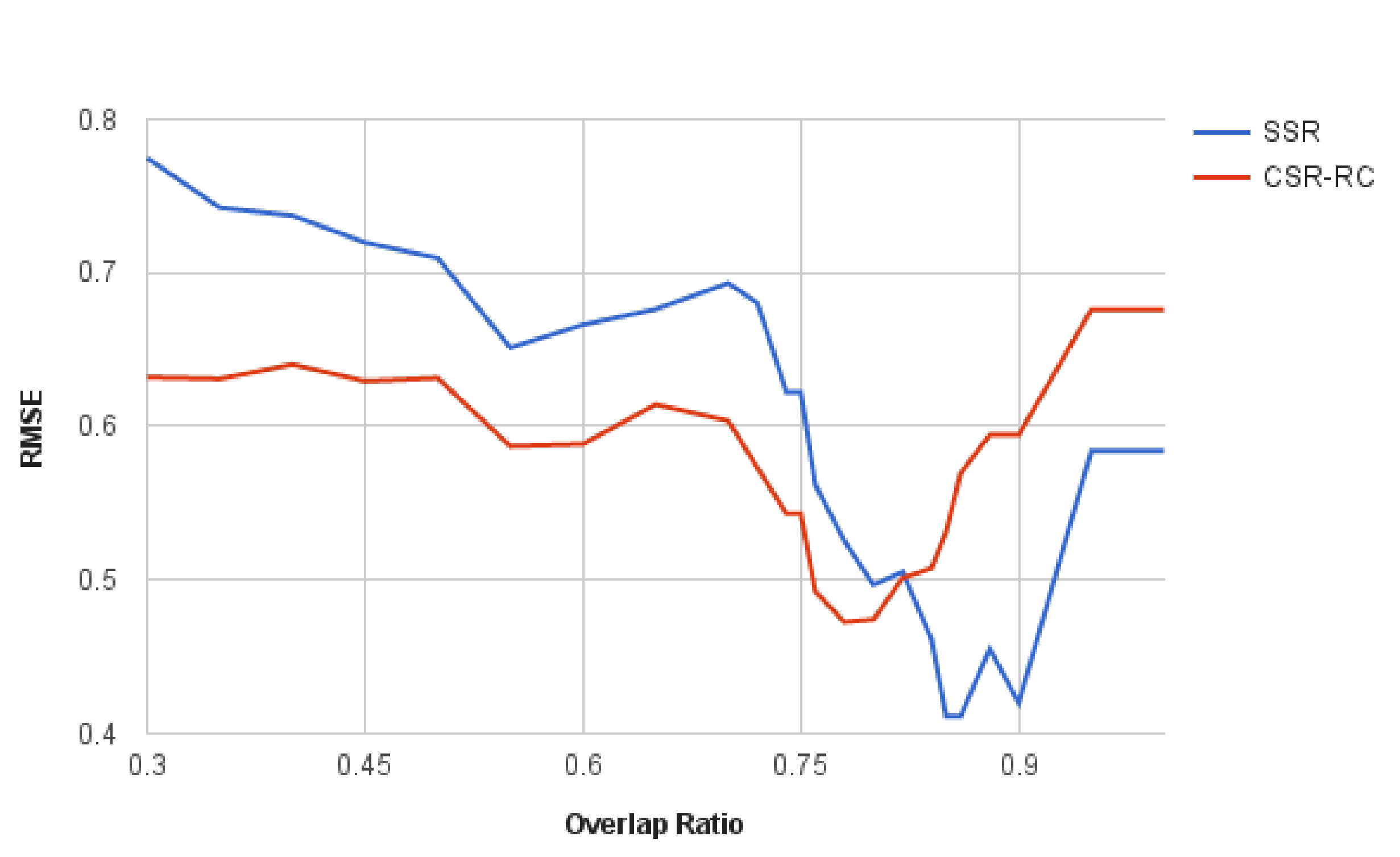}
    \caption{RMSE of SSR model compared to the CSR-RC w.r.t. overlap ratio for the case of 9 prior courses. The performance for other choices of prior courses is very similar.}
    \label{fig:3}
\end{figure}

As one of the parameters for this problem was the overlap ratio between the courses of the target student and other students, Figure~\ref{fig:2} presents the behavior of the model's RMSE (left) and AvgRMSE (right) as we vary the overlap ratio for $D^{c, \ge 5}_{test}(k=5), D^{c, \ge 7}_{test}(k=7)$ and $D^{c, \ge 9}_{test}(k=9)$. When the overlap ratio is increased, the selected students have more courses in common with the target user and that results to better performance. In order to compare the performance of SSR against CSR-RC, Figure~\ref{fig:3} shows the RMSE of the best CSR-RC and SSR models. The RMSE values were computed as the subsets of the test set that was predicted by both models. If the overlap ratio is more than 0.8, then SSR is more accurate. However, the capability of this method to predict courses is very low, i.e., we can predict 50\% less courses than the CSR model for $k=9$ when the overlap ratio is more than 0.8, because there are not as many students that had followed the same degree plan as the selected student.


\subsection{Methods based on Matrix Factorization}
\label{sec:results:mf}

The performance of the methods based on matrix factorization is shown in Table~\ref{table:6} for various number of latent factors. Besides the MF and CSMF schemes that were described in Section~\ref{sec:methods:mf}, this table also shows results for a method labeled ``MF-GB'', which is derived from the MF scheme by eliminating the global bias term ($\mu$) of Eqn.~\ref{eq:4}.
These results show that CSMF leads to lower RMSE values when there are more than nine prior courses per student, which confirms that by building matrix factorization models on smaller but denser course-specific sub-matrices, we can derive low-rank models that lead to more accurate matrix completion. Even for the case with more than five prior courses, if we focus on denser models, the majority of courses are predicted better by CSMF* than by the best model, MF-GB.
In terms of the number of latent factors, we can see that in most cases, the best performance is achieved with small number of latent factors. This should not be surprising, as the average number of grades per student is low, which does not support a large number of latent factors.
\begin{table}[bt]
\centering
\caption{The performance achieved by the methods based on matrix factorization model w.r.t. the number of prior courses and the number of latent factors.}
\setlength{\tabcolsep}{.5em}
\begin{tabular}{r|r|l| r r r r}
\toprule
Prior courses      & Latent Factors &         & MF & MF-GB     & CSMF & CSMF* \\
\midrule
\multirow{6}{*}{5}        & 2      &  \multirow{3}{*}{RMSE}    & 0.662 & 0.661 & 0.683 &0.676  \\
                   & 5           &         & 0.666 & 0.667 & 0.682 &0.682 \\
                   & 8           &         & 0.667 & 0.672 & 0.679 & 0.676  \\ \cline{2-7}
                   & 2           & \multirow{3}{*}{AvgRMSE}   & 0.597 & 0.581 & 0.648  &0.645 \\
                   & 5           & 	    & 0.603 & 0.569 & 0.643 & 0.647 \\
                   & 8           &  	    & 0.604 & 0.596 & 0.645 & 0.644 \\ \hline
\multirow{6}{*}{7}        & 2      &  \multirow{3}{*}{RMSE}    &0.667 & 0.671 & 0.684 & 0.679  \\
                   & 5           &         & 0.673 & 0.675 & 0.680 & 0.677 \\
                   & 8           &         & 0.676 & 0.681  & 0.681 & 0.676 \\\cline{2-7}
                   & 2          &  \multirow{3}{*}{AvgRMSE} & 0.590 & 0.598 & 0.641 & 0.643 \\
                   & 5           &         & 0.603 & 0.607 & 0.638 &  0.640\\
                   & 8           &         & 0.604 & 0.610 & 0.637 & 0.640 \\ \hline
\multirow{6}{*}{9}        & 2      &  \multirow{3}{*}{RMSE}    & 0.675 & 0.684 &  0.683 &0.671   \\
                   & 5           &         & 0.677 & 0.687 &  0.676 & 0.672 \\
                   & 8           &         & 0.681 & 0.692  &  0.677 & 0.674 \\ \cline{2-7} 
                   & 2           &  \multirow{3}{*}{AvgRMSE} & 0.581 & 0.600 & 0.653 &  0.648 \\
                   & 5           &         & 0.582 & 0.607 & 0.645 & 0.646 \\
                   & 8           &         & 0.579 & 0.599 & 0.648 & 0.647  \\
\bottomrule
\end{tabular}
\label{table:6}
\end{table}

\subsection {Comparison with other methods}
\label{sec:results:other}

Table~\ref{table:8} compares the performance of the baseline approaches described in Section~\ref{sec:methods:compm} (BiasOnly and SBCF) with the best-performing course-specific regression method (CSR-RC), and the best CSMF method (two latent factors). In addition, the results labeled ``CSMF$^*$'' correspond to those obtained by CSMF in which the best-performing number of latent factors for each course can be different and was selected based on their performance on the validation set (10\% of the training data).
CSR-RC and CSMF lead to RMSE and AvgRMSE values that are substantially better than either BiasOnly or SBCF. In terms of the methods that we developed, we see that CSR-RC consistently outperforms CSMF, suggesting that sparse linear regression methods are better than those based on matrix factorization for this setting. Finally, comparing the performance of CSMF$^*$ against CSMF, we see that even though the former achieved better performance, the difference is not very large, which suggests that CSMF's performance is more consistent across its different model parameters. 

\begin{table}[bt]
\centering
\caption{Comparison of the performance achieved from our methods with the competing approaches.}
\setlength{\tabcolsep}{.5em}
\begin{tabular}{l |r r r||r r r}
\toprule
 & \multicolumn{3}{c||}{RMSE}&\multicolumn{3}{c}{AvgRMSE}\\ \midrule
Prior courses & 5 & 7& 9 &  5 & 7 &9 	\\ 
\midrule
{BiasOnly}		&  		&	&0.728 	& 		& 	&0.687\\
{SBCF}		&  		&	&0.677 	& 		& 	&0.675\\
{CSR-RC}         & 0.634    &0.632 &0.632 &0.585 &0.579 &0.543 \\
{CSMF}		&0.679 	&0.680 &0.676	&0.645  	&0.638 & 0.645\\
{CSMF*}		&0.676  	&0.676 &0.671	&0.644  	&0.640 & 0.648\\
\bottomrule
\end{tabular}
\label{table:8}
\begin{tablenotes}
	\item The performance of the models trained on the different datasets were evaluated on the $D_{test}^{\ge 9}$ test set, which is the common subset among their respective test sets.
\end{tablenotes}
\end{table}

\section{Conclusions}
\label{sec:concl}

In this paper, we presented two course-specific approaches based on linear regression and matrix factorization that perform better than existing approaches based on traditional methods. This suggests that focusing on a course specific subset of the data can result in more accurate predictions. A student-course specific approach was also developed but its accuracy in grade prediction is limited by the diverse nature of degree plans. The course-specific regression was the one with the best results compared to any other method tested.

\section{Aknowledgements}
This work was supported in part by NSF (IIS-0905220, OCI-1048018, CNS-1162405, IIS-1247632, IIP-1414153, IIS-1447788) and the Digital Technology Center at the University of Minnesota. Access to research and computing facilities was provided by the Digital Technology Center and the Minnesota Supercomputing Institute. \url{http://www.msi.umn.edu}

\bibliographystyle{splncs03}
\bibliography{ref,karypis}

\end{document}